\documentclass[showpacs,twocolumn,floats]{revtex4}
\usepackage{graphicx}
\usepackage{amsmath}
\begin{document}
\title{Quantum-to-classical crossover for Andreev billiards in a magnetic field}
\author{M.\ C.\ Goorden$^1$, Ph.\ Jacquod$^2$, and C.\ W.\ J.\ Beenakker$^1$}
\affiliation{$^1$Instituut-Lorentz, Universiteit Leiden, P.O. Box 9506,
2300 RA Leiden, The Netherlands\\
$^2$D\'{e}partement de Physique Th\'{e}orique, Universit\'{e} de Gen\`{e}ve, CH-1211 Gen\`{e}ve 4, Switzerland }
\begin{abstract}
We extend the existing quasiclassical theory for the superconducting proximity effect in a chaotic quantum dot, to include a time-reversal-symmetry breaking magnetic field. Random-matrix theory (RMT) breaks down once the Ehrenfest time $\tau_E$ becomes longer than the mean time $\tau_D$ between Andreev reflections. As a consequence, the critical field at which the excitation gap closes drops below the RMT prediction as $\tau_E/\tau_D$ is increased. Our quasiclassical results are supported by comparison with a fully quantum mechanical simulation of a stroboscopic model (the Andreev kicked rotator).
\end{abstract}
\maketitle

\section{Introduction}
When a quantum dot is coupled to a superconductor via a point contact, the conversion of electron to hole excitations by Andreev reflection governs the low-energy spectrum. The density of states of such an Andreev billiard was calculated using random-matrix theory (RMT) \cite{Mel96}. If the classical dynamics in the isolated quantum dot is chaotic, a gap opens up in the spectrum. The excitation gap $E_{\rm{gap}}$ is of the order of the Thouless energy $\hbar/\tau_D$, with $\tau_D$ the average time between Andreev reflections. Although chaoticity of the dynamics is essential for the gap to open, the size of the gap in RMT is independent of the Lyapunov exponent $\lambda$ of the chaotic dynamics.

If the size $L$ of the quantum dot is much larger than the Fermi wavelength $\lambda_F$, a competing timescale $\tau_E\simeq\lambda^{-1}\ln{\left(L/\lambda_F\right)}$ appears, the Ehrenfest time, which causes the breakdown of RMT \cite{Lod98}. The gap becomes dependent on the Lyapunov exponent and for $\tau_E\gg\tau_D$ vanishes as $E_{\rm{gap}}\simeq\hbar/\tau_E$. The Ehrenfest time dependence of the gap has been investigated in several works \cite{Tar01,Ada02,Sil02,Vav03,Jac03,Goo03,Kor04}. For a recent review, see Ref.\ \cite{Bee04}.

A magnetic field breaks time-reversal symmetry, thereby reducing $E_{\rm{gap}}$. At a critical field $B_c$ the gap closes.
This was calculated using RMT in Ref.\ \cite{Mel97}, but the effect of a finite Ehrenfest time was not studied before. Here we extend the zero-field theory of Silvestrov et al. \cite{Sil02} to non-zero magnetic field. It is a quasiclassical theory, which relates the excitation spectrum to the classical dynamics in the billiard. The entire phase space is divided into two parts, depending on the time $T$ between Andreev reflections. Times $T<\tau_E$ are quantized by identifying the adiabatic invariant, while times $T>\tau_E$ are quantized by an effective RMT with $\tau_E$-dependent parameters.

There exists an alternative approach to quantization of the Andreev billiard, due to Vavilov and Larkin \cite{Vav03}, which might also be extended to non-zero magnetic field. In zero magnetic field the two models have been shown to give similar results \cite{Bee04}, so we restrict ourselves here to the approach of Ref.\ \cite{Sil02}.

The outline of the paper is as follows. We start by describing the adiabatic levels in Sec.\ \ref{sectionadiabatic}, followed by the effective RMT in Sec.\ \ref{sectioneff}. In Sec.\ \ref{sectionkick} we compare our quasiclassical theory with fully quantum mechanical computer simulations. We conclude in Sec.\ \ref{sectionconclusions}.

\section{Adiabatic quantization}\label{sectionadiabatic}

\begin{figure}
\includegraphics[width=8cm]{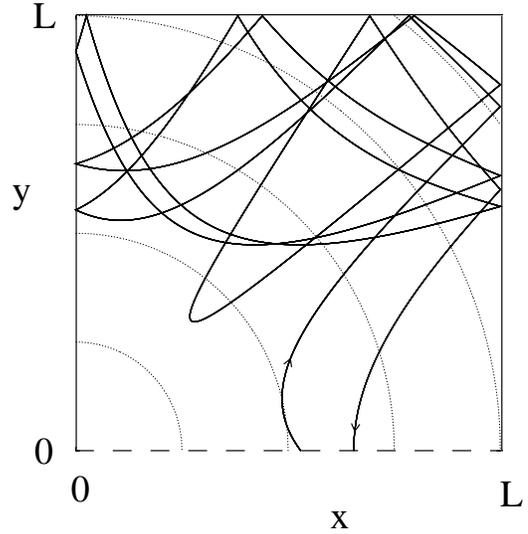}
\caption{Classical trajectory in an Andreev billiard. Particles are deflected by the potential $V=\left[(r/L)^2-1\right]V_0$ for $r<L$, $V=\left[-4(r/L)^2+10(r/L)-6\right]V_0$ for $r>L$, with $r^2=x^2+y^2$ (the dotted lines are equipotentials). At the insulating boundaries (solid lines) there is specular reflection, while the particles are Andreev reflected at the superconductor ($y=0$, dashed line). Shown is the trajectory of an electron at the Fermi level ($E=0$), for $B=0$ and $E_F=0.84\, eV_0$. The Andreev reflected hole will retrace this path.}
\label{billiard}
\end{figure}

\begin{figure}
\includegraphics[width=8cm]{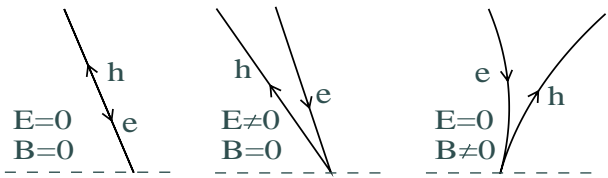}
\caption{Andreev reflection at a NS boundary (dashed line) of an electron to a hole. The left panel shows the case of perfect retroreflection (zero excitation energy $E$ and zero magnetic field $B$). The middle and right panels show that the hole does not precisely retrace the path of the electron if $E$ or $B$ are non-zero. }
\label{drift}
\end{figure}

\begin{figure}
\includegraphics[width=8cm]{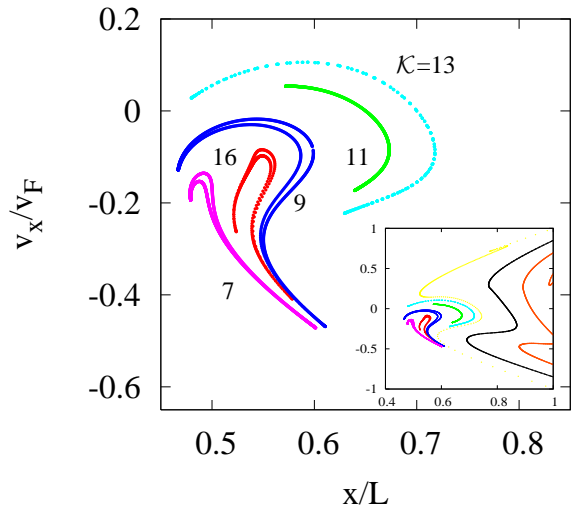}
\caption{(Color online) Poincar\'{e} map for the Andreev billiard of Fig.\ \ref{billiard}. Each dot marks the position $x$ and tangential velocity $v_x$ of an electron at the NS boundary. Subsequent dots are obtained by following the electron trajectory for $E,B\rightarrow 0$ at fixed ratio $B/E=\frac{1}{3}\sqrt{m/V_0L^2e^3}$. The inset shows the full surface of section of the Andreev billiard, while the main plot is an enlargement of the central region. The drift is along closed contours defined by ${\cal{K}}=$ constant [see Eq.\ (\ref{IKdef})]. The value of the adiabatic invariant ${\cal{K}}$ (in units of $\sqrt{mL^2/eV_0}$) is indicated for several contours. All contours are closed loops, but for some contours the opening of the loop is not visible in the figure.  }
\label{phasespace}
\end{figure}

\begin{figure}
\includegraphics[width=8cm]{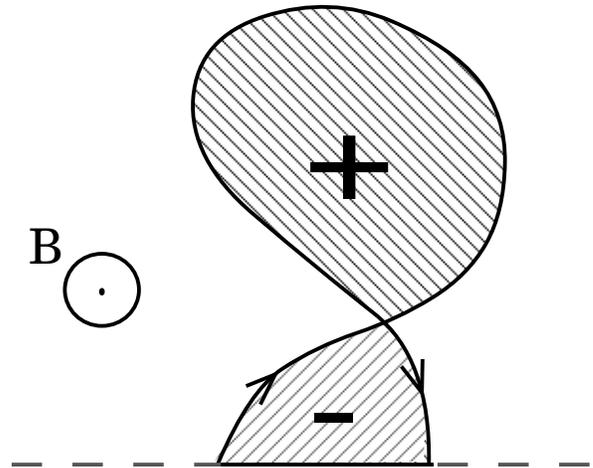}
\caption{Directed area for a classical trajectory, consisting of the area enclosed by the trajectory after joining begin and end points along the NS boundary (dashed line). Different parts of the enclosed area have different signs because the boundary is circulated in a different direction. }
\label{area}
\end{figure}

We generalize the theory of adiabatic quantization of the Andreev billiard of Ref.\ \cite{Sil02} to include the effect of a magnetic field. An example of the geometry of such a billiard is sketched in Fig.\,\ref{billiard}. The normal metal lies in the $x$-$y$ plane and the boundary with the superconductor (NS boundary) is at $y=0$. The classical mechanics of electrons and holes in such an Andreev billiard has been analyzed in Refs.\,\cite{Kos95,Wie02,Fyt05}. We first summarize the results we need, then proceed to the identification of the adiabatic invariant, and finally present its quantization.

\subsection{Classical mechanics}
The classical equation of motion
\begin{eqnarray}
\ddot{\bf{r}}(t)=-\frac{e}{m}\dot{\bf{r}}\times {\bf{B}}+\frac{e}{m}{\nabla} V({\bf{r}})
\end{eqnarray}
is the same for the electron and the hole because both charge $e$ and mass $m$ change sign. The vector ${\bf{B}}$ is the uniform magnetic field in the $z$-direction and $V({\bf r})$ is the electrostatic potential in the plane of the billiard. The dots on ${\bf{r}}=(x,y)$ denote time derivatives. We follow the classical trajectory of an electron starting at the NS boundary position $(x,0)$ with velocity $(v_x,v_y)$. The electron is at an excitation energy $E$ counted from the Fermi level. After a time $T$ the electron returns to the superconductor and is retroreflected as a hole. Retroreflection means that $v_x\rightarrow -v_x$. The $y$-component $v_y$ of the velocity also changes sign, but in addition it is slightly reduced in magnitude, $v_y^2\rightarrow v_y^2-4E/m$, so that an electron at an energy $E$ above the Fermi level becomes a hole at an energy $E$ below the Fermi level. 

This refraction is one reason why the hole does not precisely retrace the path of the electron. A second reason is that a non-zero $B$ will cause the hole trajectory to bend in the direction opposite to the electron trajectory (because the velocity has changed sign), see Fig.\ \ref{drift}. It follows that if either $E$ or $B$ are non-zero, the hole will return to the NS boundary at a slightly different position and with a slightly different velocity. The resulting drift of the quasi-periodic motion is most easily visualized in a Poincar\'{e} surface of section, see Fig.\,\ref{phasespace}. Each dot marks the position $x$ and tangential velocity $v_x$ of an electron leaving the NS boundary. At non-zero $E$ or $B$, subsequent dots are slightly displaced, tracing out a contour in the $(x,v_x)$ plane. In the limit $E,B\rightarrow 0$, the shape of these contours is determined by the adiabatic invariant of the classical dynamics. In Ref.\ \cite{Sil02} it was shown that the contours in the Poincar\'{e} surface of section are {\em isochronous} for $B=0$. This means that they are given by $T(x,v_x)=\rm{const}$, with $T(x,v_x)$ the time it takes an electron at the Fermi level to return to the NS boundary, as a function of the starting point $(x,v_x)$ on the boundary. In other words, for $B=0$ the time between Andreev reflections is an adiabatic invariant in the limit $E\rightarrow 0$. 
\subsection{Adiabatic invariant}
We generalize the construction of the adiabatic invariant of Ref.\ \cite{Sil02} to $B\neq 0$. We start from the Poincar\'{e} invariant
\begin{equation}
{\cal{I}}(t)=\oint_{C(t)}{\bf p}\cdot d{\bf r} \label{Itdef}
\end{equation}
over a closed contour $C(t)$ in phase space that moves according to the classical equations of motion. The contour extends over two sheets of phase space, joined at the NS interface. In the electron sheet the canonical momentum is ${\bf p}_{+}=m{\bf v}_{+}-e{\cal A}$, while in the hole sheet it is ${\bf p}_{-}=-m{\bf v}_{-}+e{\cal A}$. Both the velocity ${\bf v}_{\pm}$, given in absolute value by $|{\bf v}_{\pm}|=(2/m)^{1/2}[E_F\pm E+eV({\bf r})]^{1/2}$ and directed along the motion, as well as the vector potential ${\cal A}=\frac{1}{2}B\hat{z}\times{\bf r}$ are functions of the position ${\bf r}$ on the contour, determined, respectively, by the energy $E$ and the magnetic field $B$. (Since the contour is closed, the Poincar\'{e} invariant is properly gauge invariant.)

Quite generally, $d{\cal I}/dt=0$, meaning that ${\cal{I}}$ is a constant of the motion \cite{actionI}. For $E=B=0$ we take $C(0)$ to be the self-retracing orbit from electron to hole and back to electron. It is obviously time-independent, with ${\cal I}=0$ (because the contributions from electron and hole sheet cancel). For $E$ or $B$ non-zero, we construct $C(0)$ from the same closed trajectory in real space, but now with ${\bf p}_{\pm}({\bf r})$ and ${\cal A}({\bf r})$ calculated at the given values of $E$ and $B$. Consequently, this contour $C(t)$ will drift in phase space, preserving ${\cal {I}}(t)={\cal{I}}(0)$. The Poincar\'{e} invariant is of interest because it is closely related to the action integral
\begin{equation}
{I}=\oint_{O_{eh}}{\bf p}\cdot d{\bf r}.
\end{equation}
The action integral is defined as an integral along the periodic electron-hole orbit $O_{eh}$ followed by electrons and holes at $E,B=0$. To every point $(x,v_x)$ in the Poincar\'{e} surface of section corresponds an orbit $O_{eh}$ and hence an action integral $I(x,v_x)$. We compare the contour $C(t)$ and the trajectory $O_{eh}$ intersecting the Poincar\'{e} surface of section at the same point $(x,v_x)$. At $t=0$ they coincide and for sufficiently slow drifts they stay close and therefore the action integral $I={\cal{I}}(0)+{\cal{O}}(t^2)$ is an adiabatic invariant of the motion in the Poincar\'{e} surface of section \cite{actionI}.

It remains to determine the adiabatic invariant $I$ in terms of $E$ and $B$ and the chosen trajectory $C(0)$. To linear order in $E,B$ we find
\begin{equation}
I=2E{\cal{K}},\;\; {\cal{K}}\equiv T-eAB/E, \label{IKdef}
\end{equation}
with $A=\frac{1}{2}\oint ({\bf r}\times d{\bf r})\cdot\hat{z}$ the directed area (see Fig.\ \ref{area}) enclosed by the electron trajectory and the NS boundary. Both the time $T$ and the area $A$ are to be evaluated at $E=B=0$. Because $E$ is a constant of the motion, adiabatic invariance of $I$ implies that ${\cal{K}}\equiv I/2E$ is an adiabatic invariant. At zero field this adiabatic invariant is simply the time $T$ between Andreev reflections. At non-zero field the invariant time contains also an electromagnetic contribution $-eAB/E$, proportional to the enclosed flux.

Fig.\ \ref{phasespace} shows that, indeed, the drift in the Poincar\'{e} surface of section is along contours $C_{{\cal{K}}}$ of constant ${\cal{K}}$. In contrast to the zero-field case, the invariant contours in the surface of section are now no longer energy independent. This will have consequences for the quantization, as we describe next.  

\subsection{Quantization}

The two invariants $E$ and ${\cal{K}}$ define a two-dimensional torus in four-dimensional phase space. The two topologically independent closed contours on this torus are formed by the periodic electron-hole orbit $O_{eh}$ and the contour $C_{\cal{K}}$ in the Poincar\'{e} surface of section. The area they enclose is quantized following the prescription of Einstein-Brillouin-Keller \cite{Gut90,note1},
\begin{subequations}
\label{Iboth}
\begin{eqnarray}
\label{I1}
\oint_{O_{eh}} {\bf {p}}\cdot d{\bf{r}}&=&2\pi\hbar(m+1/2),\hspace{0.5cm} m=0,1,2,\hdots\\
\oint_{C_{\cal{K}}} {p_x}d{x}&=&2\pi\hbar(n+1/2),\hspace{0.5cm} n=0,1,2,\hdots
\label{I2}
\end{eqnarray}
\end{subequations}
The action integral (\ref{I1}) can be evaluated explicitly, leading to
\begin{eqnarray}
E{\cal{K}}=\pi\hbar(m+1/2).
\label{Enadiabatic}
\end{eqnarray}
The second quantization condition (\ref{I2}) gives a second relation between $E$ and ${\cal{K}}$, so that one can eliminate ${\cal{K}}$ and obtain a ladder of levels $E_{mn}$.
For $B=0$ the quantization condition (\ref{I2}) is independent of $E$, so one obtains separately a quantized time $T_n$ and quantized energy $E_{mn}=(m+1/2)\pi\hbar/T_n$. For $B\ne 0$ both ${\cal{K}}_{mn}$ and $E_{mn}$ depend on the sets of integers $m,n$.

\subsection{Lowest adiabatic level}
\begin{figure}
\includegraphics[width=8cm]{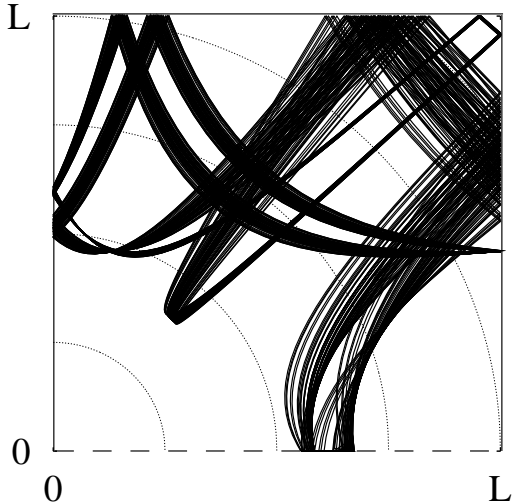}
\caption{Illustration of a bunch of trajectories within a single scattering band in the billiard defined in Fig.\ \ref{billiard}. All trajectories in this figure have starting conditions in the band containing the contour with ${\cal{K}}=11$ of Fig.\,\ref{phasespace}. Both $T$ and $A$ vary only slightly from one trajectory to the other, so that the whole band can be characterized by a single $\bar{T}$ and $\bar{A}$, being the average of $T$ and $A$ over the scattering band. }
\label{fluxtube}
\end{figure}

The value $E_{00}$ of the lowest adiabatic level follows from the pair of quantization conditions (\ref{Iboth}) with $m=n=0$. To determine this value we need to determine the area $O({\cal{K}})=\oint_{C_{\cal{K}}}p_xdx$ enclosed by contours of constant ${\cal{K}}$, in the limit of large ${\cal{K}}$. 

In Ref.\ \cite{Sil02} the area $O({\cal{K}})$ was determined in the case $B=0$, when ${\cal{K}}=T$ and the contours are isochronous. It was found that
\begin{eqnarray}
O(T)\lesssim O_0 \exp(-\lambda T),
\label{areaiso}
\end{eqnarray} 
with $\lambda$ the Lyapunov exponent of the normal billiard without superconductor and $O_0$ a characteristic area that depends on the angular distribution of the beam of electrons entering the billiard (width $L$) from the narrow contact to the superconductor (width $W$). For a collimated beam having a spread of velocities $|v_x/v_F|\lesssim W/L$ one has $O_0=Nh$. For a non-collimated beam $O_0=NhW/L$. The integer $N$ is the number of scattering channels connecting the billiard to the superconductor. The quantization requirement $O(T)\ge\pi\hbar$ gives the lowest adiabatic level in zero magnetic field \cite{Sil02},
\begin{eqnarray}
E_{00}(B=0)=\frac{\pi\hbar}{2\tau_E},\hspace{1cm} \tau_E=\frac{1}{\lambda}\ln\left(O_0/\pi\hbar\right).
\label{gapB0}
\end{eqnarray}
The Ehrenfest time $\tau_E$ corresponds to a contour that encloses an area $\pi\hbar$.

In order to generalize Eq.\ (\ref{areaiso}) to $B\ne 0$, we discuss the concept of scattering bands, introduced in Ref.\ \cite{Sil03} for a normal billiard (where they were called transmission and reflection bands). Scattering bands are ordered phase space structures that appear in open
systems, even if their closed counterparts are fully chaotic. These structures are characterized 
by regions in which
the functions $T(x,v_x)$ and $A(x,v_x)$ vary slowly almost everywhere.
Hence, they contain orbits of almost constant return time and directed area, that is,
orbits returning by bunches. One such bunch is depicted in Fig.\ \ref{fluxtube}.
The scattering bands are bounded by contours of diverging $T(x,v_x)$ and $A(x,v_x)$. The divergence is very slow ($\propto 1/\ln{\epsilon}$, with $\epsilon$ the distance from the contour \cite{Ada02}), so the mean return time $\bar{T}$ and mean directed area $\bar{A}$ in a scattering band remain finite and well defined \cite{note2}. 

The area $O_{\rm{band}}$ of a band depends on $\bar{T}$ as \cite{Sil03}
\begin{eqnarray}
\label{Oband}
O_{\rm{band}}(\bar{T})\simeq O_0\exp{(-\lambda\bar{T})}.
\end{eqnarray}
Since an isochronous contour must lie within a single scattering band, Eq.\ (\ref{areaiso}) follows from Eq.\ (\ref{Oband}) and from the fact that the distribution of return times is sharply peaked around the mean $\bar{T}$. Because contours of constant ${\cal{K}}=T-eAB/E$ must also lie within a single scattering band, the area $O({\cal{K}})$ is bounded by the same function $O_{\rm{band}}(\bar{T})$. We conclude that within a given scattering band the largest contour of constant $T$ and the largest contour of constant ${\cal{K}}$ each have approximately the same area as the band itself, 
\begin{eqnarray}
\label{areaK}
O(T), O({\cal{K}})\lesssim O_{\rm{band}}(\bar{T})\simeq O_0 \exp({-\lambda \bar{T}}).
\end{eqnarray} 

We are now ready to determine the magnetic field dependence of the lowest adiabatic level $E_{00}(B)$. The corresponding contour $C_{\cal{K}}$ lies in a band characterized by a mean return time $\bar{T}=\lambda^{-1}\ln\left(O_0/\pi\hbar\right)$, according to Eqs. (\ref{I2}) and (\ref{areaK}). This is the same Ehrenfest time as Eq.\ (\ref{gapB0}) for $B=0$ (assuming that the orbital effect of the magnetic
field does not modify $\lambda$) . The energy of the lowest adiabatic level $E_{00}$ is determined by the quantization condition (\ref{Enadiabatic}),
\begin{eqnarray}
E_{00}{\cal{K}}\approx E_{00}\tau_E+eA_{\rm{max}}B=\pi\hbar/2.
\label{quantization}
\end{eqnarray}
The range of directed areas $-A_{\rm{max}}\lesssim\bar{A}\lesssim A_{\rm{max}}$ is the product of the area $L^2$ of the billiard and the maximum number of times $n_{\rm{max}}\approx v_F\bar{T}/L$ that a trajectory can encircle that area (clockwise or counterclockwise) in a time $\bar{T}$. Hence $A_{\rm{max}}=v_F\bar{T}L\lesssim v_F\tau_EL$ and we find
\begin{eqnarray}
\label{Enadiabatic2}
E_{00}(B)\equiv E_{\rm{gap}}^{\rm{ad}}\approx\frac{\pi\hbar}{2\tau_E}-ev_FLB.
\end{eqnarray}

We conclude that a magnetic field shifts the lowest adiabatic level downward by an amount $ev_FLB$ which is independent of $\tau_E$. Eq.\ (\ref{Enadiabatic2}) holds up to a field $B_c^{\rm{ad}}$ at which the lowest adiabatic level reaches the Fermi level,
\begin{eqnarray}
\label{Bcadiabatic}
B_c^{\rm{ad}}=\frac{\pi\hbar}{2eA_{\rm{max}}}\simeq\frac{\pi\hbar}{2\tau_Eev_FL}.
\end{eqnarray}
We have added the label ``ad'', because the true critical field at which the gap closes may be smaller due to non-adiabatic levels below $E_{00}$. For $B=0$, the ground state is never an adiabatic state \cite{Bee04}. In the next section we study the effective RMT, in order to determine the contribution from non-adiabatic levels (return times $T>\tau_E$). 

\subsection{Density of states}
The pair of quantization conditions (\ref{Iboth}) determines the individual energy levels with $T<\tau_E$ and $|A|<A_{\rm{max}}=v_F\tau_EL$. 
For semiclassical systems with $L/\lambda_F\gg 1$ the level spacing $\delta$ of the isolated billiard is so small that individual levels are not resolved and it suffices to know the smoothed (or ensemble averaged) density of states $\rho_{\rm{ad}}(E)$.
In view of Eq.\ (\ref{Enadiabatic}) it is given by
\begin{eqnarray}
\rho_{\rm{ad}}(E)=N\int_0^{\tau_{E}} dT\int_{-A_{\rm{max}}}^{A_{\rm{max}}} dA\, P(T,A)\nonumber\\
\times\sum_m\delta\left(E-\frac{\pi\hbar(m+1/2)+eAB}{T}\right),
\label{rhoad}
\end{eqnarray}
in terms of the joint distribution function $P(T,A)$ of return time $T$ and directed area $A$. In the limit $\tau_E\rightarrow\infty$ this formula reduces to the Bohr-Sommerfeld quantization rule of Ref.\ \cite{Mel96} for $B=0$ and to the generalization of Ref.\ \cite{Ihr01} for $B\ne 0$.
The adiabatic density of states (\ref{rhoad}) vanishes for $E<E_{\rm{gap}}^{\rm{ad}}$. Its high energy asymptotics (meaning $E\gg E_{\rm gap}^{\rm ad}$, but still $E\ll\Delta$) can be estimated using $P(T,A)=P(A|T) P(T)$ with the conditional distribution $P(A|T)$ (which will be discussed in the next section) and the return time distribution $P(T)=\exp{(-T/\tau_D)}/\tau_D$.
One gets
\begin{equation}
 \lim_{\substack {E\rightarrow \infty\\ E\ll \Delta}}\rho_{\rm{ad}}(E)=\frac{2}{\delta}\left(1-e^{-\tau_E/\tau_D}\left[1+\frac{\tau_E}{\tau_D}\right]\right).
\label{rhoadan}
\end{equation}
The limit (\ref{rhoadan}) is less than the value $2/\delta$, which also contains the contribution from the non-adiabatic levels with $T>\tau_E$.

\section{Effective random-matrix theory}\label{sectioneff}
\begin{figure}
\includegraphics[width=8cm]{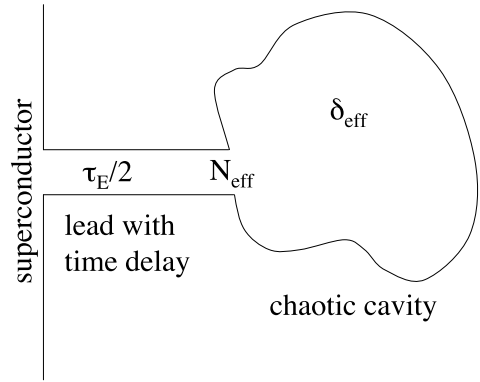}
\caption{Pictorial representation of the effective RMT of an Andreev billiard. The part of phase space with long trajectories (return time $>\tau_E$) is represented by a {\em {chaotic}} cavity with level spacing $\delta_{\rm{eff}}$, connected to the superconductor via a {\em{fictitious}} ballistic lead with $N_{\rm{eff}}$ channels. The lead introduces a channel-independent delay time $\tau_E/2$ and a channel-dependent phase shift $\phi_n$, which is different from the distribution of phase shifts in a real lead.}
\label{cavitylonglead}
\end{figure}

The adiabatic quantization applies only to the part of phase space in which the return time $T$ is less than the Ehrenfest time $\tau_E$. To quantize the remainder, with $T>\tau_E$, we apply the effective random-matrix theory (RMT) of Ref.\ \cite{Sil02}. The existing formulation \cite{Sil02,Bee04} does not yet include a magnetic field, so we begin by extending it to non-zero $B$. 

\subsection{Effective cavity}
The effective RMT is based on the decomposition of the scattering matrix in the time domain into two parts,
\begin{eqnarray}
S(t)=\left\{\begin{array}{ll}
S_{cl}(t) & \textrm{if $t<\tau_E$}\\
S_{q}(t)& \textrm{if $t>\tau_E$.}
\end{array}\right.
\end{eqnarray}
The classical, short-time part $S_{cl}(t)$ couples to $N_{cl}$ scattering channels of return time $<\tau_E$, which can be quantized adiabatically as explained in the previous section. The remaining 
\begin{equation}
N_q=N-N_{cl}=Ne^{-\tau_E/\tau_D}\equiv N_{\rm{eff}}
\end{equation}
quantum channels, with return time $>\tau_E$, are quantized by RMT with effective $\tau_E$-dependent parameters. 

To describe the effective RMT ensemble from which $S_q$ is drawn, we refer to the diagram of Fig.\ \ref{cavitylonglead}, following Ref.\ \cite{Bee04}. A wave packet of return time $t>\tau_E$ evolves along a classical trajectory for the initial $\tau_E/2$ and the final $\tau_E/2$ duration of its motion. This classical evolution is represented by a fictitious ballistic lead with delay time $\tau_E/2$, attached at one end to the superconductor. 
The transmission matrix of this lead is an $N_{\rm{eff}}\times N_{\rm{eff}}$ diagonal matrix of phase shifts $\exp\left[i\Phi(B)\right]$ (for transmission from left to right) and $\exp\left[i\Phi(-B)\right]$ (for transmission from right to left). The ballistic lead is attached at the other end to a chaotic cavity having an $N_{\rm{eff}}\times N_{\rm{eff}}$ scattering matrix $S_0$ with RMT distribution. The entire scattering matrix $S_q(t)$ of the effective cavity plus ballistic lead is, in the time domain,
\begin{eqnarray}
S_q(t)=e^{i\Phi(-B)}S_0(t-\tau_E,B)e^{i\Phi(B)},
\end{eqnarray}
and in the energy domain,
\begin{eqnarray}
S_q(E)=e^{iE\tau_E/\hbar}e^{i\Phi(-B)}S_0(E,B)e^{i\Phi(B)}.
\label{SE}
\end{eqnarray}
The level spacing $\delta_{\rm{eff}}$ of the effective cavity is increased according to 
\begin{eqnarray}
\delta_{\rm{eff}}/\delta=N/N_{\rm{eff}}=e^{\tau_E/\tau_D},
\end{eqnarray}
to ensure that the mean dwell time $2\pi\hbar/N_{\rm{eff}}\delta_{\rm{eff}}$ remains equal to $\tau_D$, independent of the Ehrenfest time.

For weak magnetic fields (such that the cyclotron radius $mv_F/eB\gg L$), the phase shifts $\Phi(B)$ are linear in $B$:
\begin{equation}
\Phi(B)\simeq\Phi(0)+B\Phi'(0)\equiv\Phi(0)+{\rm{diag}}\left[\phi_1,\phi_2\dots\phi_{N_{\rm{eff}}}\right].
\end{equation}
The phases $\phi_n$ are the channel dependent, magnetic field induced phase shifts of classical trajectories spending a time $\tau_E/2$ in a chaotic cavity. 

The conditional distribution of directed areas $A$ for a given return time $T$ is a truncated Gaussian \cite{Bar93,Ihr01},
\begin{eqnarray}
P(A|T)&\propto&\exp{\left(-A^2/A_0^2\right)}\theta(A_{\rm{max}}-|A|),\nonumber\\
A_0^2&\propto&v_FTL^3,
\label{PA}
\end{eqnarray}
with $\theta(x)$ the unit step function.
This implies that the distribution $P(\phi)$ of phase shifts $\phi=eAB/\hbar$ for $T=\tau_E/2$ is given by
\begin{eqnarray}
P(\phi)&\propto&\exp{\left[-\frac{\phi^2}{c} \frac{\tau_D}{\tau_E}\left(\frac{B_0}{B}\right)^2\right]}\theta\left(\phi_{\rm{max}}-|\phi|\right),\label{Pphi}\\
\phi_{\rm{max}}&=&\frac{eA_{\rm{max}}B}{\hbar}\simeq\frac{B}{B_0}\sqrt{\frac{v_F\tau_E^2}{L\tau_D}}.
\end{eqnarray}
The constant $c$ of order unity is determined by the billiard geometry and $B_{0}$ denotes the critical magnetic field of the Andreev billiard when $\tau_E\rightarrow 0$. Up to numerical coefficients of order unity, one has \cite{Mel97} 
\begin{equation}
B_0\simeq\frac{\hbar}{eL^2}\sqrt{\frac{L}{v_F\tau_D}}. 
\end{equation}

\subsection{Density of states}
The energy spectrum of an Andreev billiard, for energies well below the gap $\Delta$ of the bulk superconductor, is related to the scattering matrix by the determinantal equation \cite{Bee91}

\begin{eqnarray}
{\rm{Det}}\left[1+S(E)S^*(-E)\right]=0.
\label{eigenS}
\end{eqnarray}
Since $S_{cl}$ and $S_q$ couple to different channels, we may calculate separately the contribution to the spectrum from the effective cavity, governed by $S_q$. We substitute the expression (\ref{SE}) for $S_q$, to obtain  
\begin{eqnarray}
\label{deteqS}
&&{\rm{Det}}\left[1+e^{2iE\tau_E/\hbar}S_0(E,B)\Omega(B)S^*_0(-E,B)\Omega^*(B)\right]=0,\nonumber\\\\
&&\Omega(B)\equiv e^{i\Phi(B)-i\Phi(-B)}={\rm{diag}}[e^{2i\phi_1},e^{2i\phi_2}\dots e^{2i\phi_{N_{\rm{eff}}}}].\nonumber\\
\end{eqnarray}
In Ref.\ \cite{Bee04} the density of states was calculated from this equation for the case $B=0$, when $\Omega=1$. We generalize the calculation to $B\ne 0$. The technicalities are very similar to those of Ref.\ \cite{Bro97}.

The scattering matrix $S_0(E,B)$ of the open effective cavity can be represented by \cite{Bee97,Guh98}
\begin{equation}
S_0(E,B)=1-2\pi iW^{T}\left[E-H_0(B)+i\pi WW^{T}\right]^{-1}W,
\label{SvsH}
\end{equation}
in terms of the Hamiltonian $H_0(B)$ of the closed effective cavity and a coupling matrix $W$. The dimension of $H_0$ is $M\times M$ and the dimension of $W$ is $M\times N_{\rm{eff}}$. The matrix $W^{T}W$ has eigenvalues $M\delta_{\rm{eff}}/\pi^2$. The limit $M\rightarrow\infty$ at fixed level spacing $\delta_{\rm{eff}}$ is taken at the end of the calculation. Substitution of Eq.\ (\ref{SvsH}) into the determinantal equation (\ref{deteqS}) gives a conventional eigenvalue equation \cite{Bro97},
\begin{eqnarray}
&&{\rm{Det}}\left[E-H_{\rm{eff}}(B)\right]=0,\\
&&H_{\rm{eff}}(B)=\left(\begin{array}{ccc} H_0(B)&0\\0&-H_0^*(B)\end{array}\right)-\mathcal{W},\label{Heff}\\
&&\mathcal{W}=\frac{\pi}{\cos{u}}\left(\begin{array}{ccc} WW^T\sin{u}&W\Omega(B)W^{T}\\W\Omega^*(B)W^{T}&WW^{T}\sin{u}\end{array}\right).
\end{eqnarray}
We have abbreviated $u=E\tau_E/\hbar$.

The Hamiltonian $H_0(B)$ of the fictitious cavity has the Pandey-Mehta distribution \cite{Meh91},
\begin{eqnarray}
P(H)&\propto&\exp\biggl(-\frac{\pi^2(1+b^2)}{4M\delta_{\rm{eff}}^2}\nonumber\\
&&\times\sum_{i,j=1}^M\left[({\rm{Re}}H_{ij})^2+b^{-2}({\rm{Im}} H_{ij})^2\right]\biggr).
\label{PandeyMehta}
\end{eqnarray}
The parameter $b\in[0,1]$ measures the strength of the time-reversal symmetry breaking. It is related to the magnetic field by \cite{Mel97}
\begin{equation}
\label{bvsB}
\frac{M}{N_{\rm{eff}}}b^2=\frac{1}{8}(B/B_0)^2.
\end{equation}

The ensemble averaged density of states $\rho_{\rm{eff}}(E)$ is obtained from the Green function,
\begin{eqnarray}
\rho_{\rm{eff}}(E)&=&-\frac{1}{\pi} {\rm{Im}} {\rm{Tr}} \left(1+\frac{d\mathcal{W}}{dE}\right)\mathcal{G}(E+i0^+),\label{rho}\\
\mathcal{G}(z)&=&\langle(z-H_{\rm{eff}})^{-1}\rangle,
\end{eqnarray}
where the average $\langle\cdots\rangle$  is taken with the distribution (\ref{PandeyMehta}).
Using the results of Refs.\ \cite{Mel97,Bro97} we obtain a self-consistency equation for the trace of the ensemble averaged Green function, 
\begin{eqnarray}
G=\left(\begin{array}{ccc} G_{11}&G_{12}\\G_{21}&G_{22}\end{array}\right)=\frac{\delta}{\pi}\left(\begin{array}{ccc} {\rm Tr}\, \mathcal{G}_{11}&{\rm Tr}\, \mathcal{G}_{12}\\{\rm Tr}\, \mathcal{G}_{21}&{\rm Tr}\, \mathcal{G}_{22}\end{array}\right).
\end{eqnarray}
The four blocks refer to the block decomposition (\ref{Heff}) of the effective Hamiltonian.
The self-consistency equation reads
\begin{widetext}
\begin{eqnarray}
&&G_{11}=G_{22}, \hspace{1cm}G_{12}G_{21}=1+G_{11}^2,
\label{Gsum1}\\
&&0=N_{\rm{eff}}\left(\frac{E}{2E_T}-\left(\frac{B}{B_0}\right)^2 \frac{G_{11}}{2}\right)G_{12}+\sum_{j=1}^{N_{\rm{eff}}}\frac{e^{2i\phi_j}G_{11}+G_{12}\sin{u}}{\frac{1}{2}\left[e^{-2i\phi_j}G_{12}+e^{2i\phi_j}G_{21}\right]+\cos{u}+G_{11}\sin{u}},
\label{Gsum2}\\
&&0=N_{\rm{eff}}\left(\frac{E}{2E_T}-\left(\frac{B}{B_0}\right)^2 \frac{G_{11}}{2}\right)G_{21}+\sum_{j=1}^{N_{\rm{eff}}}\frac{e^{-2i\phi_j}G_{11}+G_{21}\sin{u}}{\frac{1}{2}\left[e^{-2i\phi_j}G_{12}+e^{2i\phi_j}G_{21}\right]+\cos{u}+G_{11}\sin{u}},
\label{Gsum3}
\end{eqnarray}
with the Thouless energy $E_T=\hbar/2\tau_D$.

From Eq.\ (\ref{rho}) we find the density of states

\begin{eqnarray}
&&\rho_{\rm{eff}}(E)=-\frac{2}{\delta_{\rm{eff}}}{\rm{Im}}\left[G_{11}+\frac{\tau_E}{\tau_D\cos{u}}\sum_{j=1}^{N_{\rm{eff}}}\frac{G_{11}+\frac{1}{2}\sin{u}\left(G_{21}e^{2i\phi_j}+G_{12}e^{-2i\phi_j}\right)}{\cos{u}+G_{11}\sin{u}+\frac{1}{2}G_{12}e^{-2i\phi_j}+\frac{1}{2}G_{21}e^{2i\phi_j}}\right].
\label{rhosum}
\end{eqnarray}
\end{widetext}
Because $N_{\rm{eff}}\gg 1$, we may replace in Eqs. (\ref{Gsum1}--\ref{rhosum}) the sum $\sum_jf(\phi_j)$ by $\int d\phi\,P(\phi)$, with $P(\phi)$ given by Eq.\ (\ref{Pphi}). In the next section we will compare the density of states obtained from (\ref{Gsum1}--\ref{rhosum}) with a fully quantum mechanical calculation. In this section we discuss the low and high energy asymptotics of the density of states.


In the limit $E\rightarrow\infty$, $E\ll\Delta$ we find from Eqs. (\ref{Gsum1}--\ref{Gsum3}) that $G_{12}=G_{21}\propto 1/E\rightarrow 0$ while $G_{11}\rightarrow -i$. Substituting this into Eq.\ (\ref{rhosum}) we obtain the high energy limit,
\begin{eqnarray}
\lim_{\substack {E\rightarrow \infty\\ E\ll\Delta}}\rho_{\rm{eff}}(E)&=&\frac{2}{\delta_{\rm{eff}}}\left(1+\frac{\tau_E}{\tau_D}\right)\nonumber\\
&=&\frac{2}{\delta}e^{-\tau_E/\tau_D}\left(1+\frac{\tau_E}{\tau_D}\right).
\label{rhoeffhighE}
\end{eqnarray}
This limit is larger than $2/\delta_{\rm{eff}}$ because of the contribution from states in the lead, cf. Fig.\ \ref{cavitylonglead}.
Together with Eq.\ (\ref{rhoadan}) we find that the total density of states,
\begin{eqnarray}
\rho(E)=\rho_{\rm{eff}}(E)+\rho_{\rm{ad}}(E),
\label{rhotot}
\end{eqnarray}
tends to $2/\delta$ for high energies, as it should be.

At low energies the density of states $\rho_{\rm{eff}}(E)$ obtained from the effective RMT vanishes for $E<E_{\rm{gap}}^{\rm{eff}}$. In the limit $\tau_E\gg\tau_D$ the lowest level in the effective cavity is determined by the fictitious lead with return time $\tau_E$. This gives the same gap as for adiabatic quantization,
\begin{eqnarray}
E_{\rm{gap}}^{\rm{eff}}=E_{\rm{gap}}^{\rm{ad}}=\frac{\hbar}{\tau_E}\left(\frac{\pi}{2}-2\phi_{\rm{max}}\right)\approx\frac{\pi\hbar}{2\tau_E}-ev_FLB,
\label{Enefflarge}
\end{eqnarray}
cf. Eq.\ (\ref{Enadiabatic2}).
The two critical magnetic fields $B_c^{\rm{eff}}$ and $B_c^{\rm{ad}}$ coincide in this limit,
\begin{equation}
B_c^{\rm{eff}}=B_c^{\rm{ad}}\approx\frac{\pi\hbar}{2\tau_Eev_FL}\simeq B_0\sqrt{\frac{\tau_DL}{v_F\tau_E^2}}, \,\mbox{if $\tau_E\gg\tau_D$},
\end{equation}
cf. Eq.\ (\ref{Bcadiabatic}).
In the opposite regime of small $\tau_E$ we find a critical field of
\begin{eqnarray}
\label{Bceffsmall}
B_c^{\rm{eff}}=B_0\left(1-\frac{c\tau_E}{8\tau_D}\right),\,\mbox{if $\tau_E\ll\sqrt{L\tau_D/v_F}$},
\end{eqnarray}
which is smaller than $B_c^{\rm{ad}}$ so $B_c=B_c^{\rm{eff}}$.
In the intermediate regime $\sqrt{L\tau_D/v_F}\lesssim\tau_E\lesssim\tau_D$, the critical field $B_c$ is given by
\begin{equation}
B_c={\rm min}\left(B_c^{\rm{eff}},B_c^{\rm{ad}}\right).
\label{Bcmin}
\end{equation}
We do not have an analytical formula for $B_c^{\rm{eff}}$ in this intermediate regime, but we will show in the next section that $B_c^{\rm{ad}}$ drops below $B_c^{\rm{eff}}$ so that $B_c=B_c^{\rm{ad}}$.

\section{Comparison with quantum mechanical model}\label{sectionkick}
In this section we compare our quasiclassical theory with a quantum mechanical model of the Andreev billiard. The model we use is the Andreev kicked rotator introduced in Ref.\ \cite{Jac03}.
We include the magnetic field into the model using the three-kick representation of Ref.\ \cite{Two04}, to break time-reversal-symmetry at both the quantum mechanical and the classical level. The basic equations of the model are summarized in Appendix \ref{app.b}.

\begin{figure*}
\includegraphics[width=16cm]{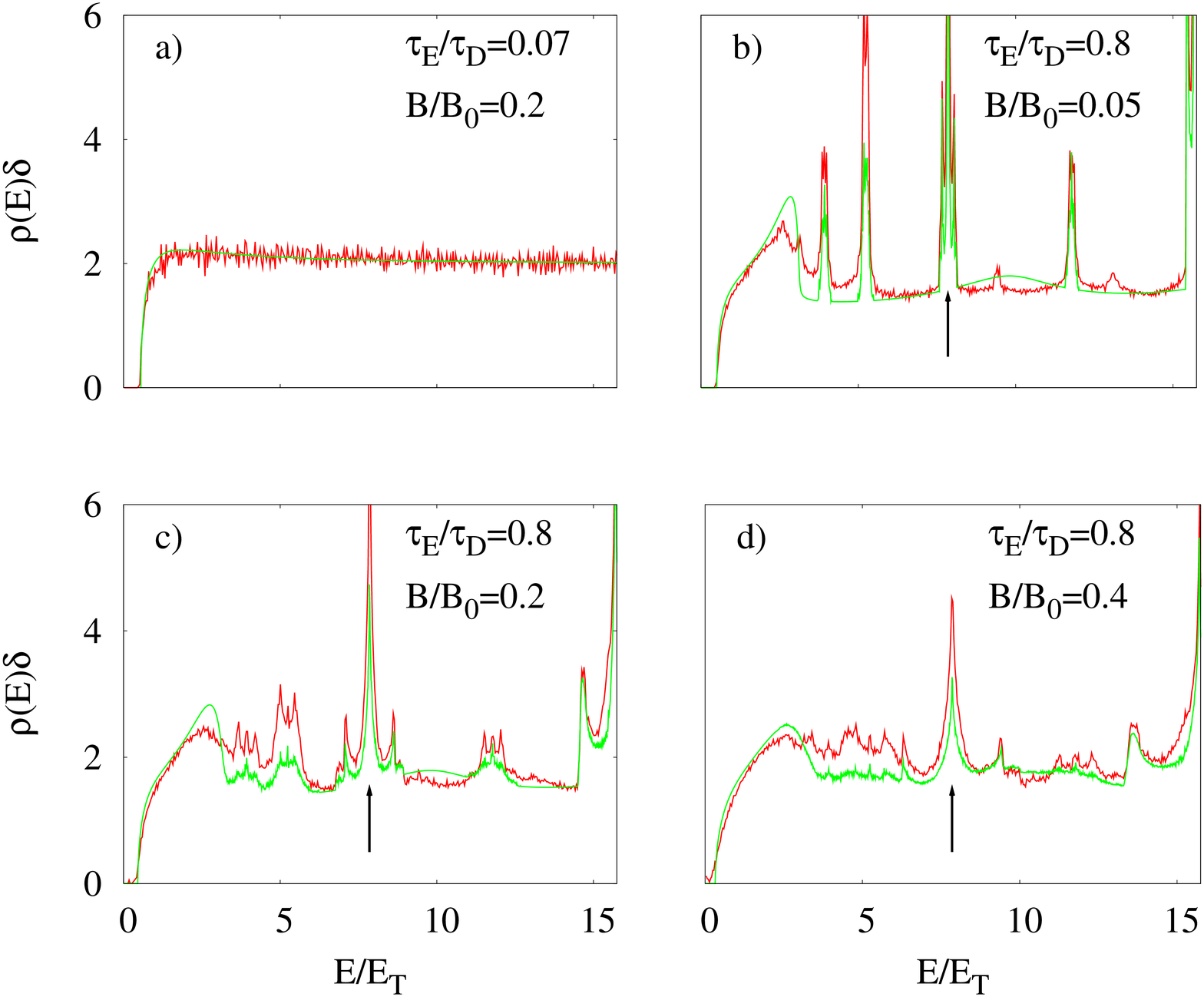}
\caption{(Color online) Ensemble averaged density of states $\rho(E)$ of the Andreev kicked rotator. The dark (red) curves show the numerical results from the fully quantum mechanical model, while the light (green) curves are obtained from Eq.\ (\ref{rhotot}) with input from the classical limit of the model. The energy is scaled by the Thouless energy $E_T=\hbar/2\tau_D$ and the density is scaled by the level spacing $\delta$ of the isolated billiard. The parameters of the kicked rotator are $M=2048$, $N=204$, $q=0.2$, $K=200$ in panel a and $M=16384$, $N=3246$, $q=0.2$, $K=14$ in panels b, c, d. The three-peak structure indicated by the arrow in panels b, c, d is explained in Fig.\ \ref{PAfig}.}
\label{rho3kick}
\end{figure*}

\begin{figure}
\includegraphics[width=8cm]{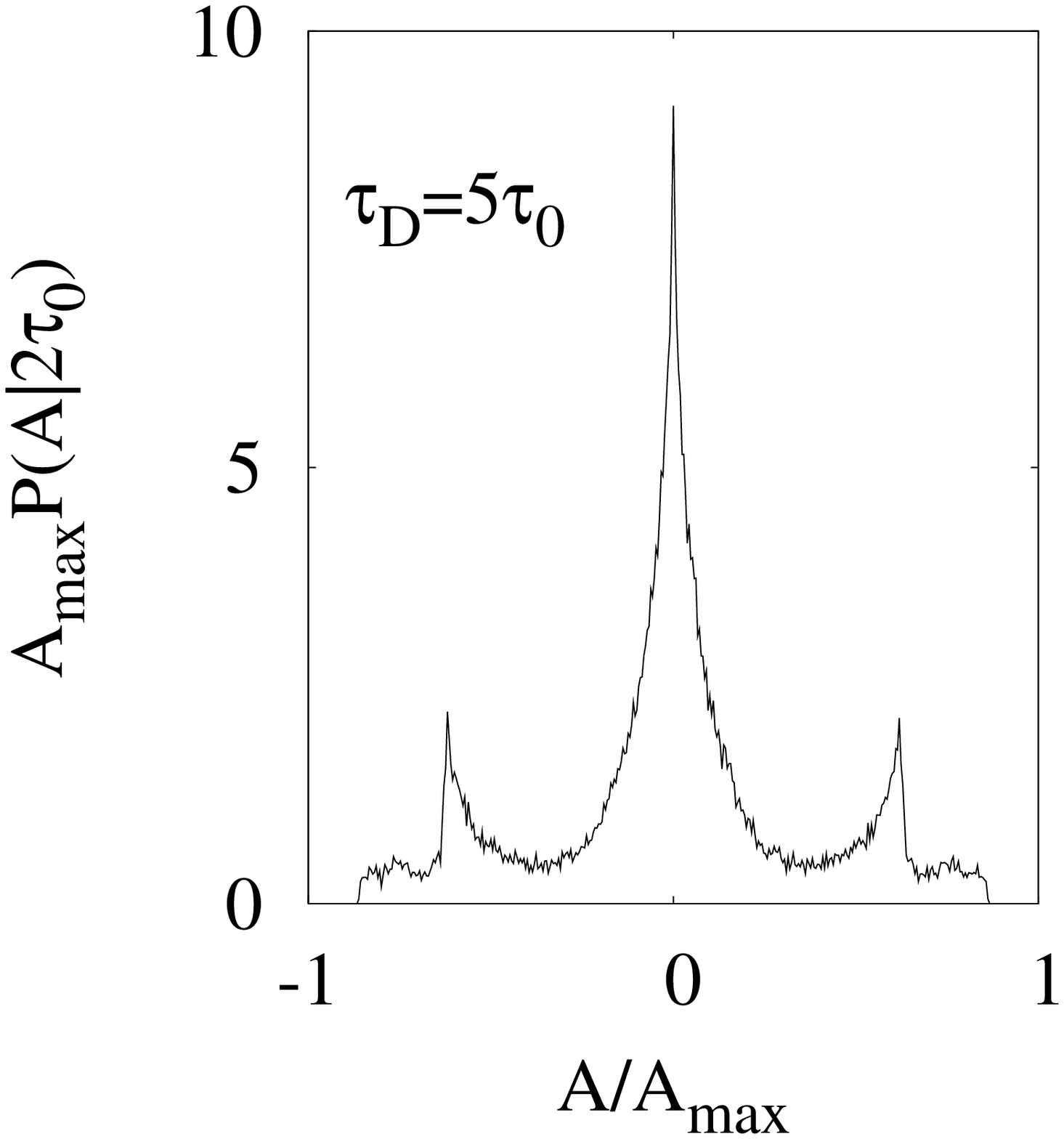}
\caption{Conditional distribution $P(A|T)$ of directed areas $A$ enclosed by classical trajectories with $T=2\tau_0$, for $K=14$, $q=0.2$ and $\tau_D=5\tau_0$. The distribution was obtained from the classical map (\ref{classicalmap}) at $\gamma=0$. Trajectories with $T=2\tau_0$ give rise to a peak in the density of states centered around $E/E_T=(m+1/2)\pi\hbar/2\tau_0$, cf. Eq. (\ref{rhoad}). On the energy scale of Fig.\ \ref{rho3kick} only the peak with $m=0$ can be seen, at $E/E_T=2.5\,\pi\approx 7.9$. In a magnetic field this peak broadens and it obtains the side peaks of $P(A|2\tau_0)$.
}
\label{PAfig}
\end{figure}

In Fig.\ \ref{rho3kick} we show the ensemble averaged density of states of the Andreev kicked rotator and we compare it with the theoretical result (\ref{rhotot}). The Ehrenfest time is given by \cite{Vav03,Jac03}
\begin{eqnarray}
\label{taueorder1}
\tau_E=\lambda^{-1}\left[\ln{(N^2/M)}+{\cal{O}}(1)\right],
\end{eqnarray}
with $M$ the dimensionality of the Floquet matrix. We neglect the correction term of order unity. The mean dwell time is $\tau_D=(M/N)\tau_0$ and the level spacing is $\delta=(2\pi/M)\hbar/\tau_0$, with $\tau_0$ the stroboscopic time.
The relation between $B/B_0$ and the parameters of the kicked rotator is given by Eq.\ (\ref{Bvsgamma}).

In Fig.\ \ref{rho3kick}a $\tau_E\ll\tau_D$ and we recover the RMT result of Ref.\ \cite{Mel97}. The density of states is featureless with a shallow maximum just above the gap. In Figs.\ \ref{rho3kick}b, c,
d $\tau_E$ and $\tau_D$ are comparable. Now the spectrum consists of both adiabatic levels (return time $T<\tau_E$) as well as effective RMT levels (return time $T>\tau_E$). The adiabatic levels cluster in peaks, while the effective RMT forms the smooth background, with a pronounced bump above the gap.

The peaks in the excitation spectrum of the Andreev kicked rotator appear because the return time $T$ in Eq.\ (\ref{rhoad}) is a multiple of the stroboscopic time $\tau_0$ \cite{Jac03}. The peaks are broadened by the magnetic field and they acquire side peaks, due to the structure of the area distribution $P(A|T)$ for $T$ a small multiple of $\tau_0$. This is illustrated in Fig.\ \ref{PAfig} for the central peak of Fig.\ \ref{rho3kick}. The distribution was calculated from the classical map (\ref{classicalmap}) associated with the quantum kicked rotator. The same map gave the coefficient $c=0.55$ appearing in Eq.\ (\ref{Pphi}).

\begin{figure}
\includegraphics[width=8cm]{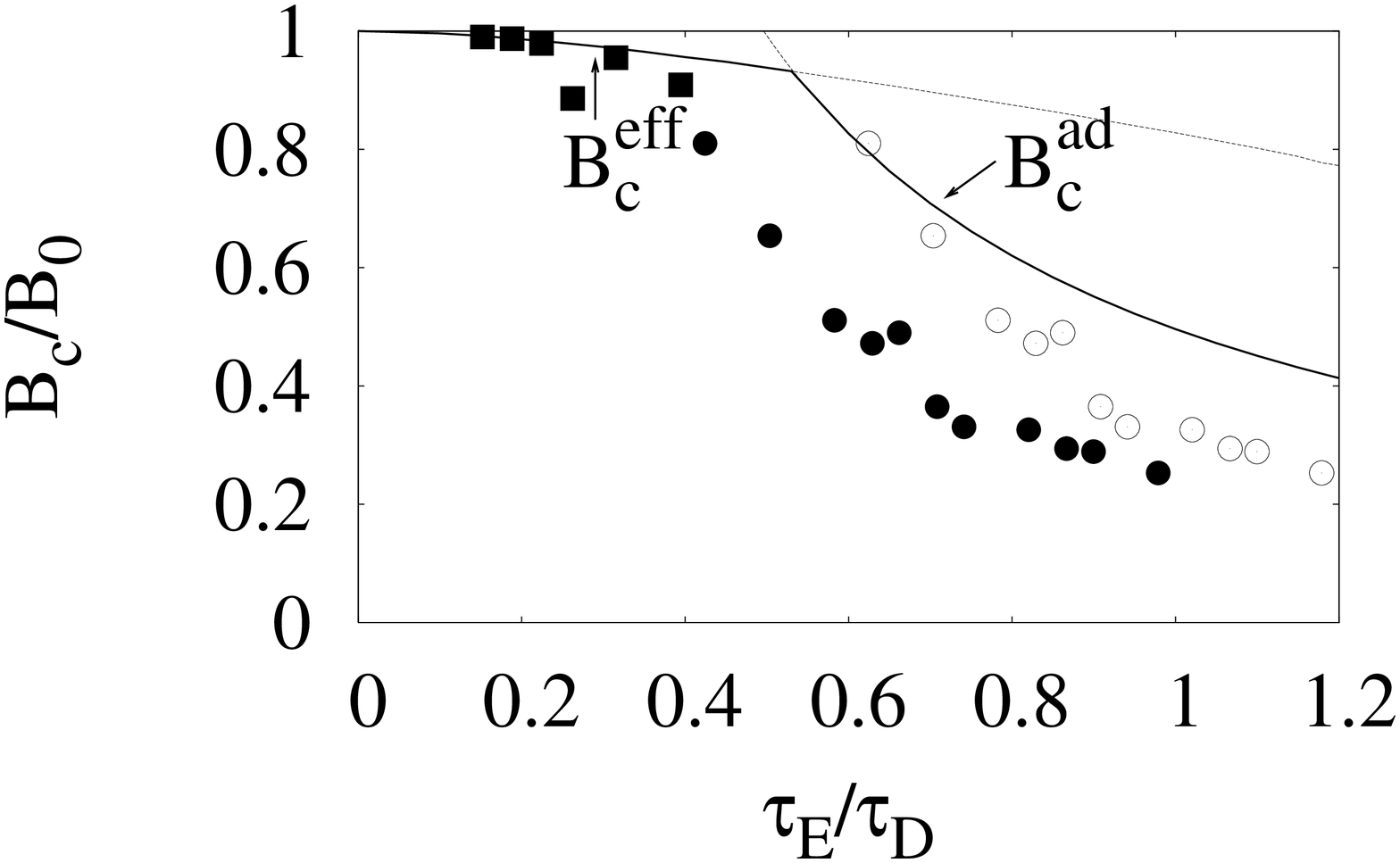}
\caption{Critical magnetic field $B_c$ of the Andreev kicked rotator as a function of the Ehrenfest time. The Ehrenfest time $\tau_E=\lambda^{-1}\ln(N^2/M)$ is changed by varying $M$ and $N$ while keeping $q=0.2$ and $\tau_D/\tau_0=M/N=5$ constant. For the closed circles the kicking strength $K=14$, while for the squares from left to right $K=4000$, $1000$, $400$, $200$, $100$, $50$. The solid curve is the quasiclassical prediction (\ref{Bcmin}). The open circles are obtained from the closed circles by the transformation $\lambda\tau_E\rightarrow\lambda\tau_E+1.75$, allowed by the terms of order unity in Eq.\ (\ref{taueorder1}).}
\label{gammacclas}
\end{figure}

In Fig.\ \ref{gammacclas} we have plotted the critical magnetic field $B_c$ at which the gap closes, as a function of the Ehrenfest time. For $\tau_E\ll\tau_D$ the Andreev kicked rotator gives a value for $B_c$ close to the prediction $B_0$ of RMT, cf. Eq.\ (\ref{Bvsgamma}). With increasing $\tau_E$ we find that $B_c$ decreases quite strongly. In the figure we also show the critical magnetic fields $B_c^{\rm{ad}}$ for adiabatic levels and $B_c^{\rm{eff}}$ for effective RMT. The former follows from Eqs.\ (\ref{Bcadiabatic}) and (\ref{Amax}),
\begin{equation}
B_c^{\rm{ad}}=\frac{\pi}{4}B_0\sqrt{\frac{2\tau_D\tau_0}{\tau_E^2}},
\end{equation}
and the latter from solving Eqs.\ (\ref{Gsum1}--\ref{Gsum3}) numerically. As already announced in the previous section, $B_c^{\rm{ad}}$ drops below $B_c^{\rm{eff}}$ with increasing $\tau_E$, which means that the lowest level $E_{\rm{gap}}$ is an adiabatic level corresponding to a return time $T<\tau_E$. 
The critical magnetic field is the smallest value of $B_c^{\rm{eff}}$ and $B_c^{\rm{ad}}$, as indicated by the solid curve. The data of the Andreev kicked rotator follows the trend of the quasiclassical theory, although quite substantial discrepancies remain. Our quasiclassical theory seems to overestimate the lowest adiabatic level, which also causes deviations between theory and numerical data in the low energy behaviour of the density of states (cf. Fig.\ \ref{rho3kick}, panels c, d).
Part of these discrepancies can be attributed to the correction term of order unity in Eq. (\ref{taueorder1}), as shown by the open circles in Fig.\ \ref{gammacclas}.

\begin{figure}
\includegraphics[width=8cm]{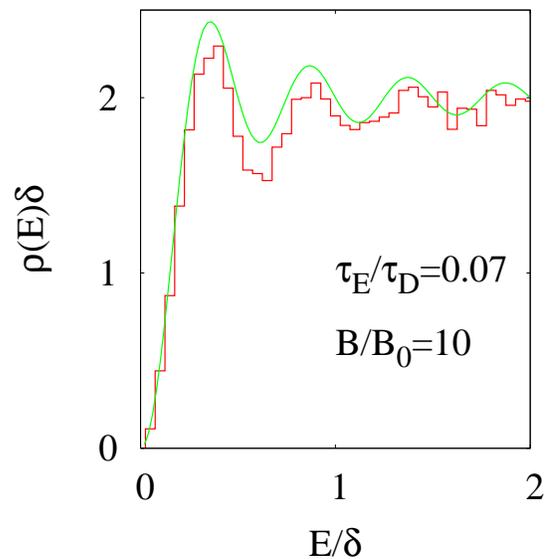}
\caption{(Color online) Ensemble averaged density of states of the Andreev kicked rotator for fully broken TRS. The histogram shows the numerical results, while the curve is the theoretical prediction (\ref{rhoLUE}) of the Laguerre unitary ensemble. Both the energy and the density of states are scaled by the level spacing $\delta$ of the isolated billiard.  The parameters of the kicked rotator are $M=2048$, $N=204$, $q=0.2$, while $K$ was varied between $200$ and $250$ to obtain an ensemble average. }
\label{LUE}
\end{figure}

In the regime of fully broken time-reversal-symmetry the distribution of eigenvalues is determined by the Laguerre unitary ensemble of RMT \cite{Alt96,Fra96}.
The ensemble averaged density of states vanishes quadratically near zero energy, according to
\begin{equation}
\rho(E)=\frac{2}{\delta}\left(1-\frac{\sin{(4\pi E/\delta)}}{4\pi E/\delta}\right).
\label{rhoLUE}
\end{equation}
In Fig.\ \ref{LUE} we show the results for the Andreev kicked rotator in this regime and we find a good agreement with Eq.\ (\ref{rhoLUE}) for $\tau_E\ll\tau_D$. We did not investigate the $\tau_E$ dependence in this regime.

\section{Conclusion}\label{sectionconclusions}
We have calculated the excitation spectrum of an Andreev billiard in a magnetic field, both using a quasiclassical and a fully quantum mechanical approach. The quasiclassical theory needs as input the classical distribution of times $T$ between Andreev reflections and directed areas $A$ enclosed in that time $T$. Times $T$ smaller than the Ehrenfest time $\tau_E$ are quantized via the adiabatic invariant and times $T>\tau_E$ are quantized by an effective random-matrix theory with $\tau_E$-dependent parameters. This separation of phase space into two parts, introduced in Ref.\ \cite{Sil02}, has received much theoretical support in the context of transport \cite{Sil03,Two03,Two04b,Two04,Jac04,Two04ph,Whi04}. The present work shows that it can be successfully used to describe the consequences of time-reversal symmetry breaking on the superconducting proximity effect.

The adiabatically quantized and effective RMT spectra each have an excitation gap which closes at different magnetic fields. The critical magnetic field $B_c$ of the Andreev billiard is the smallest of the two values $B_c^{\rm{ad}}$ and $B_c^{\rm{eff}}$. 
For relatively small Ehrenfest time $\tau_E\ll\tau_D$ the critical field $B_c^{\rm{eff}}$ from effective RMT is smaller than the critical field $B_c^{\rm{ad}}$ of the adiabatic levels, so $B_c=B_c^{\rm{eff}}$. This value $B_c^{\rm{eff}}$ is smaller than the value $B_0$ of conventional RMT \cite{Mel97}, because of the $\tau_E$-dependence of the parameters in effective RMT.
For $\tau_E\gg\tau_D$ the two fields $B_c^{\rm{ad}}$ and $B_c^{\rm{eff}}$ coincide, but in an intermediate regime of comparable $\tau_E$ and $\tau_D$ the adiabatic value $B_c^{\rm{ad}}$ drops below the effective RMT value $B_c^{\rm{eff}}$. This is indeed what we have found in the specific model that we have investigated, the Andreev kicked rotator \cite{Jac03}. The lowest level has $T<\tau_E$ for sufficiently large $\tau_E$ and $B$. This is a novel feature of the Andreev billiard in a magnetic field: For unbroken time-reversal symmetry the lowest level always corresponds to longer trajectories $T>\tau_E$ \cite{Goo03}, and thus cannot be obtained by adiabatic quantization \cite{Sil02,Bee04}.

\section*{Acknowledgements}
We have benefitted from discussions with P.\ Silvestrov and J.\ Tworzyd{\l}o.
This work was supported by the Research Training Network of the European Union on ``Fundamentals of Nanoelectronics'', by the Dutch Science Foundation NWO/FOM and by the Swiss National Science Foundation.

\appendix
\section{Andreev kicked rotator in a magnetic field}\label{app.b}
The Andreev kicked rotator in zero magnetic field was introduced in Ref.\ \cite{Jac03}. Here we give the extension to non-zero magnetic field used in Sec.\ \ref{sectionkick}.
We start from the kicked rotator with broken time-reversal symmetry but without the superconductor. The kicked rotator provides a stroboscopic description of scattering inside a quantum dot. The propagation of a state from time $t$ to time $t+\tau_0$ is given by the $M\times M$ unitary Floquet operator $F$ with matrix elements \cite{Two04}
\begin{equation}
F_{mn}=\left(X\Pi Y^*\Pi Y\Pi X\right)_{mn}.
\label{Fkick}
\end{equation}
The three matrices $X$, $Y$, and $\Pi$ are defined by
\begin{eqnarray}
Y_{mn}&=&\delta_{mn}e^{i(M\gamma/6\pi)\cos{(2\pi m/M)}},\\
X_{mn}&=&\delta_{mn}e^{-i(M/12\pi)V(2\pi m/M)},\\
\Pi_{mn}&=&M^{-1/2}e^{-i\pi/4}\exp\left[{i(\pi/M)(m-n)^2}\right].
\end{eqnarray}
The potential 
\begin{equation}
V(\theta)=K\cos{(\pi q/2)}\cos{(\theta)}+\frac{K}{2}\sin{(\pi q/2)}\sin{(2\theta)}
\end{equation}
breaks the parity symmetry for $q\ne 0$. Time-reversal symmetry is broken by the parameter $\gamma$. For kicking strengths $K\gtrsim 7$ the classical dynamics of the kicked rotator is chaotic. 

The Floquet operator (\ref{Fkick}) describes electron excitations above the Fermi level.
The hole excitations below the Fermi level are described by the Floquet operator $F^*$. Electrons and holes are coupled by Andreev reflection at the superconductor.
The $N\times M$ matrix $P$, with elements
\begin{eqnarray}
P_{nm}=\delta_{nm}\times\left\{\begin{array}{ccc}1 \mbox{ if $L_0\le n\le L_0+N-1$}\\
0\mbox{ otherwise}\end{array}\right.,
\end{eqnarray}
projects onto the contact with the superconductor. The integer $L_0$ indicates the location of the contact and $N$ is its width, in units of $\lambda_F/2$. We will perform ensemble averages by varying $L_0$. The process of Andreev reflection is described by the $2M\times 2M$ matrix
\begin{eqnarray}
{\cal{P}}=\left(\begin{array}{ccc}1-P^TP&&-iP^TP\\-iP^TP&&1-P^TP\end{array}\right).
\end{eqnarray}
The Floquet operator for the Andreev kicked rotator is constructed from the two matrices $F$ and ${\cal{P}}$ \cite{Jac03},
\begin{eqnarray}
{\cal{F}}={\cal{P}}^{1/2}\left(\begin{array}{ccc}F&&0\\0&&F^*\end{array}\right){\cal{P}}^{1/2}.
\label{Ftotal}
\end{eqnarray}
The $2M\times 2M$ unitary matrix $\cal{F}$ can be diagonalized efficiently using the Lanczos technique in combination with the fast-Fourier-transform algorithm \cite{Ket99}. The eigenvalues $e^{i\varepsilon_m}$ define the quasi-energies $\varepsilon_m\in[0,2\pi]$. One gap is centered around $\varepsilon=0$ and another gap around $\varepsilon=\pi$. For $N\ll M$ the two gaps are decoupled and we can study the gap around $\varepsilon=0$ by itself.

The correspondence between the TRS-breaking parameter $\gamma$ of the kicked rotator and the Pandey-Mehta parameter $b$ for $K\gg 1$ is given by \cite{Two04}
\begin{eqnarray}
\lim_{K\rightarrow\infty} b\sqrt{M_H}=\frac{\gamma M^{3/2}}{12\pi}.
\end{eqnarray}
Here $M_H$ is the size of the Pandey-Mehta Hamiltonian \cite{Meh91}.
Comparison with Eq.\ (\ref{bvsB}) gives the relation between $\gamma$ and the magnetic field $B$,
\begin{eqnarray}
\frac{M^{3/2}}{N^{1/2}}\gamma=\sqrt{\frac{\tau_D}{\tau_0}}M\gamma=3\pi\sqrt{2}\frac{B}{B_0}.
\label{Bvsgamma}
\end{eqnarray}
In RMT the gap closes when $B=B_0$, so when $\gamma=\gamma_0=3\pi M^{-1}\sqrt{2\tau_0/\tau_D}$.

For the quasiclassical theory we need the classical map associated with the Floquet operator (\ref{Ftotal}). The classical phase space consists of the torus $0\le\theta\le 2\pi$, $0\le p\le 6\pi$. The classical map is described by a set of equations that map initial coordinates $(\theta,p)$ onto final coordinates $(\theta',p')$ after one period $\tau_0$ \cite{Two04},
\begin{eqnarray}
\theta_1&=&\theta\pm p/3-V'(\theta)/6-2\pi\sigma_{\theta_1},\nonumber\\
p_1&=&p\mp\gamma\sin{(\theta_1)}\mp V'(\theta)/2-6\pi\sigma_{p_1},\nonumber\\
\theta_2&=&\theta_1\pm p_1/3-2\pi\sigma_{\theta_2},\nonumber\\
p_2&=&p_1-6\pi\sigma_{p_2},\nonumber\\
\theta'&=&\theta_2\pm p_2/3+\gamma\sin{(\theta_2)}/3-2\pi\sigma_{\theta'},\nonumber\\
p'&=&p_2\pm\gamma\sin(\theta_2)\mp V'(\theta')/2-6\pi\sigma_p'.
\label{classicalmap}
\end{eqnarray}
The upper/lower signs correspond to electron/hole dynamics and $V'(\theta)=dV/d\theta$.
The integers $\sigma_\theta$ and $\sigma_p$ are the winding numbers of a trajectory on the torus.

The directed area enclosed by a classical trajectory between Andreev reflections can be calculated from the difference in classical action between two trajectories related by TRS, one with $\gamma=0$ and one with infinitesimal $\gamma$. To linear order in $\gamma$ the action difference $\Delta S$ acquired after one period is given by \cite{Two04}
\begin{eqnarray}
\Delta S=\gamma\left(\cos{\theta_1}-\cos{\theta_2}\right).
\label{action}
\end{eqnarray}
The effective Planck constant of the kicked rotator is $\hbar_{\rm{eff}}=6\pi/M$, so we may obtain the increment in directed area $\Delta A$ corresponding to $\Delta S$ from
\begin{equation}
\frac{e}{\hbar}B\Delta A=\frac{\Delta S}{\hbar_{\rm{eff}}}=\frac{M}{6\pi}\gamma\left(\cos\theta_1-\cos\theta_2\right).
\end{equation}
Since $|\cos{\theta_1}-\cos{\theta_2}|<2$, the maximum directed area $A_{\rm{max}}$ acquired after $T/\tau_0$ periods is
\begin{eqnarray}
A_{\rm{max}}=2\frac{T}{\tau_0}\frac{\hbar}{eB_0}\sqrt\frac{\tau_0}{2\tau_D}.
\label{Amax}
\end{eqnarray}


\begin{thebibliography}{99}
\bibitem{Mel96} J.\ A.\ Melsen, P.\ W.\ Brouwer, K.\ M.\ Frahm, and C.\ W.\ J.\ Beenakker, Europhys.\ Lett.\ {\bf 35}, 7 (1996).
\bibitem{Lod98} A.\ Lodder and Yu.\ V.\ Nazarov, Phys.\ Rev.\ B {\bf{58}}, 5783 (1998).
\bibitem{Tar01} D.\ Taras-Semchuk and A.\ Altland, Phys.\ Rev.\ B {\bf 64}, 014512 (2001).
\bibitem{Ada02} \.{I}.\ Adagideli and C.\ W.\ J.\ Beenakker, Phys.\ Rev.\ Lett.\ {\bf{89}}, 237002 (2002).
\bibitem{Sil02} P.\ G.\ Silvestrov, M.\ C.\ Goorden, and C.\ W.\ J.\ Beenakker, Phys.\ Rev.\ Lett.\ {\bf 90}, 116801 (2003).
\bibitem{Vav03} M.\ G.\ Vavilov and A.\ I.\ Larkin, Phys.\ Rev.\ B {\bf 67}, 115335 (2003).
\bibitem{Jac03} Ph.\ Jacquod, H.\ Schomerus, and C.\ W.\ J.\ Beenakker, Phys.\ Rev.\ Lett.\ {\bf 90}, 207004 (2003).
\bibitem{Goo03} M.\ C.\ Goorden, Ph.\ Jacquod, and C.\ W.\ J.\ Beenakker, Phys.\ Rev.\ B {\bf 68}, 220501(R) (2003).
\bibitem{Kor04} A.\ Korm\'{a}nyos, Z.\ Kaufmann, C.\ J.\ Lambert, and J.\ Cserti, Phys.\ Rev.\ B {\bf 70}, 052512 (2004).
\bibitem{Bee04} C.\ W.\ J.\ Beenakker, Lect. Notes Phys. {\bf 667}, 131 (2005);  cond-mat/0406018.
\bibitem{Mel97} J.\ A.\ Melsen, P.\ W.\ Brouwer, K.\ M.\ Frahm, and C.\ W.\ J.\ Beenakker, Physica Scripta {\bf 69}, 223 (1997).
\bibitem{Kos95} I.\ Kosztin, D.\ L.\ Maslov, and P.\ M.\ Goldbart, Phys.\ Rev.\ Lett.\ {\bf 75}, 1735 (1995).
\bibitem{Wie02} J.\ Wiersig, Phys.\ Rev.\ E {\bf 65}, 036221 (2002).
\bibitem{Fyt05} N.\ G.\ Fytas, F.\ K.\ Diakonos, P.\ Schmelcher, M.\ Scheid, A.\ Lassl, K.\ Richter, and G.\ Fagas, cond-mat/0504322.
\bibitem{actionI} J.\ V.\ Jos\'{e}\ and E.\ J.\ Saletan, {\em Classical Dynamics} (Cambridge University Press, Cambridge, 1998).
\bibitem{Gut90} M.\ C.\ Gutzwiller, {\em Chaos in Classical and Quantum Mechanics} (Springer, Berlin, 1990).
\bibitem{note1} The shift by $1/2$ in Eqs. (\ref{I1}) and (\ref{I2}) accounts for two phase shifts of $\pi/2$ incurred at each Andreev reflection and at each turning point, respectively; turning points do not contribute a net phase shift to Eq. (\ref{I1}) because the phase shifts in the electron and hole sheets cancel.
\bibitem{Sil03} P.\ G.\ Silvestrov, M.\ C.\ Goorden, and C.\ W.\ J.\ Beenakker, Phys.\ Rev.\ B {\bf{67}}, 241301(R) (2003).
 \bibitem{note2} In Ref.\ \cite{Sil03} the fluctuations of $T$ around $\bar{T}$ within a single scattering band were estimated at $\delta T\simeq W/v_F\ll\tau_D$, and similarly we estimate that $\delta A\simeq WL\ll v_F\tau_DL$.
\bibitem{Ihr01} W.\ Ihra, M.\ Leadbeater, J.\ L.\ Vega, and K.\ Richter, Eur.\ Phys.\ J.\ B  {\bf 21}, 425 (2001).
\bibitem{Bar93} H.\ U.\ Baranger, R.\ A.\ Jalabert, and A.\ D.\ Stone, Chaos {\bf{3}}, 665 (1993).
\bibitem{Bee91} C.\ W.\ J.\ Beenakker, Phys.\ Rev.\ Lett.\ {\bf{67}}, 3836 (1991).
\bibitem{Bro97} P.\ W.\ Brouwer and C.\ W.\ J.\ Beenakker, Chaos, Solitons and Fractals {\bf 8}, 1249 (1997). 
\bibitem{Guh98} T.\ Guhr, A.\ M\" {u}ller-Groeling, and H.\ A.\ Weidenm\" {u}ller, Phys.\ Rep.\ {\bf{299}}, 189 (1998). 
\bibitem{Bee97} C.\ W.\ J.\ Beenakker, Rev.\ Mod.\ Phys.\ {\bf{69}}, 731 (1997).
\bibitem{Meh91} M.\ L.\ Mehta, {\em Random Matrices} (Academic, New York, 1991).
\bibitem{Two04} J.\ Tworzyd{\l}o, A.\ Tajic, and C.\ W.\ J.\ Beenakker, Phys.\ Rev.\ B {\bf 70}, 205324 (2004).
\bibitem{Alt96} A.\ Altland and M.\ R.\ Zirnbauer, Phys.\ Rev.\ Lett.\ {\bf 76}, 3420 (1996).
\bibitem{Fra96} K.\ M.\ Frahm, P.\ W.\ Brouwer, J.\ A.\ Melsen, and C.\ W.\ J.\ Beenakker, Phys.\ Rev.\ Lett.\ {\bf{76}}, 2981 (1996).
\bibitem{Two03} J.\ Tworzyd{\l}o, A.\ Tajic, H.\ Schomerus, and C.\ W.\ J.\ Beenakker, Phys.\ Rev.\ B {\bf 68}  115313 (2003)
\bibitem{Two04b}  J.\ Tworzyd{\l}o, A.\ Tajic, and C.\ W.\ J.\ Beenakker, Phys.\ Rev.\ B {\bf 69}, 165318 (2004).
\bibitem{Jac04} Ph.\ Jacquod and E.\ V.\ Sukhorukov, Phys.\ Rev.\ Lett.\ {\bf{92}}, 116801 (2004).
\bibitem{Two04ph} J.\ Tworzyd{\l}o, A.\ Tajic, H.\ Schomerus, P.\ W.\ Brouwer, and C.\ W.\ J.\ Beenakker, Phys.\ Rev.\ Lett.\ {\bf{93}}, 186806 (2004).
\bibitem{Whi04} R.\ S.\ Whitney and Ph.\ Jacquod, Phys.\ Rev.\ Lett.\ {\bf 94}, 116801 (2005).
\bibitem{Ket99} R.\ Ketzmerick, K.\ Kruse, and T.\ Geisel, Physica D {\bf {131}}, 247 (1999).







\end{thebibliography}
\end{document}